\documentclass[fleqn,usenatbib]{mnras}

\usepackage{newtxtext,newtxmath}
\usepackage[T1]{fontenc}
\usepackage{ae,aecompl}

\usepackage{graphicx}	% Including figure files
\usepackage{amsmath,multirow}	% Advanced maths commands
\usepackage{amssymb}	% Extra maths symbols
\usepackage{mathtools}  % <-- added
\usepackage{makecell}   % <-- added
\usepackage{cases}
\usepackage[dvipsnames]{xcolor}
\usepackage{adjustbox}
\definecolor{orange}{rgb}{1,0.5,0}

\newcommand{\simba}{{\sc Simba}}

\newcommand{\lya}{Ly$\alpha$}
\newcommand{\HI}{\ion{H}{i}}

\newcommand{\nhi}{N_{\rm HI}}

\newcommand{\hmpc}{h^{-1}{\rm Mpc}}

\newcommand{\ghi}{\Gamma_{\rm HI}}
\newcommand{\kms}{\;{\rm km}\,{\rm s}^{-1}}

\newcommand\cdunits{{\rm cm}^{-2}}

\title[DLAs from Simba vs TD]{Testing Galaxy Formation Simulations with Damped Lyman-$\alpha$ Abundance and Metallicity Evolution}

\author[Hassan et al.]{
Sultan Hassan$^{1,2}$\thanks{E-mail: shassan@nmsu.edu}\thanks{Tombaugh Fellow}
Kristian Finlator,$^{1,3}$
Romeel Dav\'e,$^{4,2,5}$
Christopher W. Churchill,$^{1}$ \newauthor
and J. Xavier Prochaska$^{6,7}$
\\ \\
$^{1}$ Department of Astronomy, New Mexico State University, Las Cruces, NM 88003, USA \\
$^{2}$ Department of Physics and Astronomy, University of the Western Cape, Bellville, Cape Town 7535, South Africa \\
$^{3}$ Cosmic Dawn Center (DAWN), Niels Bohr Institute, University of Copenhagen / DTU-Space, Technical University of Denmark\\
$^{4}$ Institute for Astronomy, Royal Observatory, Edinburgh EH9 3HJ, UK \\
$^{5}$ South African Astronomical Observatory, Observatory, Cape Town 7925, South Africa  \\
$^{6}$ Department of Astronomy and Astrophysics, UCO/Lick Observatory, University of
California, 1156 High Street, Santa Cruz, CA 95064\\
$^{7}$ Kavli Institute for the Physics and Mathematics of the Universe (Kavli IPMU),
5-1-5 Kashiwanoha, Kashiwa, 277-8583, Japan\\
}

\date{Accepted XXX. Received YYY; in original form ZZZ}

\pubyear{2019}
\begin{document}
\label{firstpage}
\pagerange{\pageref{firstpage}--\pageref{lastpage}}
\maketitle

\begin{abstract}
We examine the properties of damped Lyman-$\alpha$ absorbers (DLAs) emerging from a single set of cosmological initial conditions in two state-of-the-art cosmological hydrodynamic simulations: {\sc Simba} and {\sc Technicolor Dawn}.  The former includes star formation and black hole feedback treatments that yield a good match with low-redshift galaxy properties, while the latter uses multi-frequency radiative transfer to model an inhomogeneous ultraviolet background (UVB) self-consistently and is calibrated to match the Thomson scattering optical depth, UVB amplitude, and Ly-$\alpha$ forest mean transmission at $z>5$. Both simulations are in reasonable agreement with the measured stellar mass and star formation rate functions at $z\geq 3$, and both reproduce the  observed neutral hydrogen cosmological mass density, $\Omega_{\rm HI}(z)$.  However, the DLA abundance and metallicity distribution are sensitive to the   galactic outflows' feedback and the UVB amplitude. Adopting a strong UVB and/or slow outflows under-produces the observed DLA abundance, but yields broad agreement with the observed DLA metallicity distribution. By contrast, faster outflows eject metals to larger distances, yielding more metal-rich DLAs whose observational selection may be more sensitive to dust bias. The DLA metallicity distribution in models adopting an ${\rm H}_2$-regulated star formation recipe includes a tail extending to $[M/H] \ll -3$, lower than any DLA observed to date, owing to curtailed star formation in low-metallicity galaxies.  Our results show that DLA observations play an imporant role in constraining key physical ingredients in  galaxy formation models, complementing traditional ensemble statistics such as the stellar mass and star formation rate functions.
\end{abstract}


%%%%%%%%%%%%%%%%%%%%%%%%%%%%%%%%%%%%%%%%%%%%%%%%%%

%%%%%%%%%%%%%%%%% BODY OF PAPER %%%%%%%%%%%%%%%%%%

\section{Introduction}
Strong \HI\ absorbers with column densities above 2$\times$10$^{20}$ cm$^{-2}$, known as Damped Lyman-$\alpha$ systems~\citep[DLAs,][]{Wolfe:1986,Wolfe:2005,Prochaska:2005,Prochaska:2009}, are rare and distinct profiles in the spectrum of background quasars. 
DLAs contain the dominant reservoir of cosmic neutral gas that eventually feeds star formation within galaxies, and hence they provide a unique way to constrain star formation and associated processes in galaxy formation models. 

The nature of DLAs and their relationship to host galaxies have long been a source of controversy. Early works proposed that DLAs are the progenitors of present-day galaxies~\citep[e.g.][]{Lanzetta:1991,Wolfe:1995}, and \citet{Prochaska:1997} favored a scenario where DLAs are thick, rapidly-rotating disks, owing in part to their large internal kinematic velocities, now confirmed in larger samples~\citep{Neeleman:2013}.  Such large disks at early epochs were a challenge to favored cold dark matter (CDM) cosmologies~\citep{Prochaska:1998}.  
Subsequently, clustering studies have broadly constrained the DLA hosting halo mass scale to be 10$^{9-11}$ M$_{\odot}$ \citep[e.g.][]{Jeon:2019,Perez-Rafols:2018,Lochhaas:2016, Pontzen:2008}, and virial velocity range of about 50 $-$ 200 km/s~\citep[.e.g][]{Barnes:2009,Bird:2014}.  This suggested that DLAs occur in a wide range of galaxies and environments.  Observations of DLA metallicities showed that they had relatively low metallicities compared to typical star-forming galaxies at the same epoch along with alpha element ratios comparable to halo stars, with mild evolution and a clear metallicity floor ~\citep[e.g.][]{Rafelski:2012,Rafelski:2014}. These results provide constraints on the nature of DLAs, but attempts to comprehensively fit all these observations within a single cosmologically-situated model remain elusive.

The dynamic nature of neutral hydrogen gas within hierarchically growing galaxies favors the use of hydrodynamic simulations to interpret DLA observations.  Early 
simulations by \citet{Haehnelt:1998} countered the notion that DLAs challenge CDM models by
showing that irregular protogalactic clumps can equivalently reproduce the observed distribution of DLA velocity widths within a hierarchical context
\citep[but see][]{Prochaska:2010}. 
\citet{Nagamine:2004} showed that cosmological simulations can approximately reproduce the abundance of DLAs over cosmic time when star formation feedback is included.  However, \citet{Berry:2014} used a semi-analytic model to show that it is possible to match DLA abundances, but only in a model where the \HI\ disk is highly extended compared to expectations, and such models still fail at $z>3$.

More recent simulations have shown that DLA abundances are particularly sensitive to assumptions regarding stellar feedback.  \citet{Bird:2014} used cosmological simulations with {\sc Arepo} to demonstrate a degeneracy between the wind speed and UVB amplitude in which both were anti-correlated with the DLA cross section. \citet{Bird:2015} further showed that hierarchical models can statistically reproduce DLA kinematics.  \citet{Faucher:2015} used higher-resolution Feedback in Realistic Environments (FIRE) zoom simulations to show that stellar feedback has a large impact on the cross-section of high-column \HI\ gas.  \citet{Rhodin:2019} used similarly high-resolution simulations to explore the immpact parameter distribution of DLAs from host galaxies at lower redshifts, showing that high resolution and efficient star formation feedback are required to match observations.
These studies indicate that DLA properties such as their abundances, metallicities, impact parameters, and kinematic spreads provide constraints on processes of star formation and the strength of the local photo-ionising flux.

To make progress in understanding DLAs in a cosmological context, ideally one requires simulations with sufficient volume for good statistics, sufficient resolution to fully capture relevant dense gas processes, a model for star formation and associated feedback that is concordant with a wide range of observations, and self-consistent modeling of the photo-ionising background including self-shielding.  Unfortunately, no simulation currently exists that fully satisfies all these criteria. Nonetheless, recent models have made substantial progress towards this, and by comparing simulations with contrasting implementations of these processes, it is possible to gain more robust insights into the nature of DLAs within hierarchical structure formation models.  This is the goal of this paper, the first in a series to explore $z\geq 3$ DLA properties in two state of the art high-resolution cosmological simulations.

In this work, we use the observed DLA abundance and metallicity evolution to test two simulations, namely, {\sc Simba}~\citep{Dave:2019}, a cosmological hydrodynamic simulation with black hole growth feedback; and {\sc Technicolor Dawn}~\citep[TD;][]{Finlator:2018}, a cosmological hydrodynamic simulations including an on-the-fly multi-frequency radiative transfer solver. While the ultraviolet ionizing background (UVB) treatment, feedback effects and star formation recipes are all quite different, both simulations have been calibrated to reproduce key galaxy observables. For instance, {\sc Simba} reproduces~\citep{Dave:2019} galaxy stellar mass function from $z=0-6$, the stellar mass--star formation rate main sequence, low-$z$ \HI\ and $H_2$ mass fractions, the mass-metallicity relation at $z\sim 0-2.5$, black hole--galaxy co-evolution~\citep{Thomas:2019}, and galaxy dust properties~\citep{Li:2019}.  {\sc TD}, which focuses more on high redshifts, reproduces the history of reionization, the galaxy stellar mass--star formation rate relation, 
the abundance of high-$z$ metal absorbers, the ultraviolet background (UVB) amplitude, and the Lyman-$\alpha$ flux power spectrum at $z = 5.4$.  Hence both simulations appears to have sub-grid recipes for star formation and other processes that yield agreement with a range of current constraints.

The need for complementary tests of these codes arises from the fact that, although they both reproduce a broad variety of observations of galaxy growth, they do so by modeling star formation and feedback in different ways. {\sc TD} adopts the~\citet{Springel:2003} sub-grid multi-phase model in which the local star formation rate depends only on gas density, whereas {\sc Simba} employs an H$_{2}-$regulated star formation model based on the sub-grid model of~\citet{Krumholz:2009}.  While both employ kinetic star formation feedback, {\sc Simba} employs scalings taken from FIRE~\citep{Angles:2017a} that have significantly lower outflow speeds but higher mass loading than those assumed in TD which come from tuned constraints~\citep{Dave:2013}. Finally, {\sc Simba} employs a different and potentially improved hydrodynamics solver relative to that in TD.  These differences are expected to impact the predicted circum-galactic medium (CGM) neutral gas reservoirs, and thus DLAs provide complementary, sensitive tests that provide novel constraints on models.  Here, we employ the exact same initial conditions for our {\sc Simba} and TD runs down to $z=3$, therefore allowing us to isolate the differences in DLA predictions purely owing to input physics variations.

This paper is organised as follows.  We describe the {\sc Simba} and TD simulations in \S\ref{sec:sim}. In \S\ref{sec:dlamock} we cast sightlines through the simulation volumes and extract catalogs of synthetic DLAs along with their associated low and high ion transitions, including a comparison between the results of identifying DLAs in observed versus theoretical spaces. We compare the predicted gas density distribution, star formation history, and UVB treatments in \S\ref{sec:simcom}.   We explore predictions for DLA abundance evolution and the column density distribution in \S\ref{sec:dlaabund}, and metallicity distribution and evolution in \S\ref{sec:dlamet}. We summarise our results in \S\ref{sec:conc}.

\section{Simulations}\label{sec:sim}

We here briefly describe {\sc Simba} and {\sc TD} and refer the reader to~\citet{Dave:2019,Dave:2016} and~\citet{Finlator:2018} for more detailed information of the physics implemented in these simulations.

\subsection{\sc Simba}
The {\sc Simba} model was introduced in ~\citet{Dave:2019}. {\sc Simba} is a follow-up to the {\sc Mufasa}~\citep{Dave:2016} cosmological galaxy formation simulation using {\sc Gizmo}'s meshless finite mass hydrodynamics solver~\citep{Hopkins:2015,Hopkins:2018}. Radiative cooling and photo-ionisation heating are implemented using the updated {\sc Grackle-3.1} library~\citep{Smith:2017}. A spatially-uniform ionizing background from~\citet{Haardt:2012} is assumed, in which self-shielding is accounted for following~\citet{Rahmati:2013}. The chemical enrichment model tracks nine elements (C,N,O,Ne,Mg,Si,S,Ca,Fe) arising from Type II supernovae (SNe), Type Ia SNe, and Asymptotic Giant Branch (AGB) stars. Type Ia SNe and AGB wind heating are also included. The star formation-driven galactic winds are kinetically launched and decoupled into hot and cold phase winds, and  the winds are metal-loaded owing to local enrichment from supernovae, with overall metal mass being conserved. The mass rate entering galactic outflows is modelled with a broken power law following~\citet{Angles:2017b}.  The quasi-linear scaling of wind velocity with escape velocity from~\citet{Muratov:2015} is adopted. Exact equations are summarized in Table~\ref{tab:sims}. {\sc Simba} further implements black hole growth via a torque-limited accretion model~\citep{Angles:2017a}, which is a unique feature of {\sc Simba}, plus Bondi accretion from hot gas. Feedback from Active Galactic Nuclei (AGN) is implemented in two modes, a radiative mode at high accretion rates that follows constraints from observed ionised AGN  outflows~\citep[e.g.][]{Perna:2017a,Fabian:2012}, and a jet mode at low accretion rates that ejects material at velocities approaching $10^4\kms$. {\sc Simba} also accounts for the X-ray AGN feedback in the surrounding gas following~\citet{Choi:2012}. Dust production and destruction are modeled on-the-fly, leading to predictions of dust abundance that broadly agree with a variety of constraints over cosmic time~\citep{Li:2019}.

Given the importance of the star formation recipe in setting the gas content in the ISM, we describe this in somewhat more detail.
{\sc Simba} employs a molecular gas-based prescription following~\citet[hereafter KMT]{Krumholz:2009} to form stars. KMT is a physically motivated recipe to model star formation as seen in local disk galaxies, where a strong correlation is seen between SFR and molecular gas content~\citep[e.g.][]{Leroy:2008}. In the KMT model, the H$_{2}$ mass fraction $f_{\rm H_{2}}$ depends on the metallicity and local column density as follows:
\begin{equation}
  f_{\rm H_{2}} = 1 -  0.75 \frac{s}{1 + 0.25 s} \, , 
\end{equation}
where
\begin{equation}
    s = \frac{\ln{(1 + 0.6 \chi + 0.01 \chi^{2})}}{0.0396\,Z\,(\Sigma/ M_{\odot} pc^{-2})}\, ,
\end{equation}
where $Z$ is the metallicity in solar units, $\chi$ is a function of metallicity given in KMT, and $\Sigma$ is the column density.  A resolution-varying clumping factor is also implemented. This improvement over the constant value in the original KMT model, which was calibrated on $\sim$kpc scale, enables higher-resolution calculations to adopt a higher threshold density for $H_2$ formation.  Finally, ISM gas is pressurised to keep the Jeans mass resolved at all densities, resulting in an equation of state where $T\propto \rho^{1/3}$ at the highest densities.

Because the KMT model predicts a steep dependence of star formation efficiency on metallicity and was calibrated on a Milky Way-like ISM, the extrapolation to low-metallicity situations is highly uncertain. As such, {\sc Simba} implements a metallicity floor for $Z$; that is, we set $Z={\rm MAX}(Z,Z_{\rm floor})$ in the KMT equations.  Note that this does not impact the overall metallicity of the gas or the metal cooling rate; it is only a floor applied when using the KMT formulae.  Importantly, this adjustment allows star formation to occur in primordial gas, where a literal adoption of the KMT model would prevent star formation outright because metals are required for molecules to form.  The fiducial value is  $\log\ Z_{\rm floor}=-2$, but we will evaluate how sensitive DLA properties are to this choice by comparing models in which $\log\ Z_{\rm floor}=[-1,-2,-3]$.

Given $f_{\rm H_{2}}$, the SFR of an individual gas element is given by a~\citet{Schmidt:1959} Law as follows:
\begin{equation}
    \frac{d M_{*}}{dt} = \epsilon_{*} \frac{\rho f_{\rm H_{2}}}{t_{\rm dyn}}\,
\end{equation}
where $\rho$ is the gas density, $t_{\rm dyn} = 1/\sqrt{G \rho}$ is the local dynamical time, and the star formation efficiency $\epsilon_{*}$ is set to 0.02~\citep{Kennicutt:1998}. Star formation is only allowed in the dense gas phase (ISM gas) above a hydrogen number density $n_{\rm H}$ $\geq$ 0.13 cm$^{-3}$, though in practice the H$_2$ fraction forces star formation to occur at much higher densities ($n_{\rm H}\gg 1$~cm$^{-3}$). {\sc Simba} assumes a ~\citet{Chabrier:2003} initial mass function (IMF) throughout.

\subsection{\sc Technicolor Dawn (TD)}
The {\sc TD} simulations~\citep{Finlator:2018}, are a suite of cosmological radiative hydrodynamic transfer simulations. Hydrodynamics are modeled using a density-independent formulation of Smoothed Particle Hydrodynamics~\citep[SPH;][]{Hopkins:2013}. Radiative cooling is implemented following~\citet{Katz:1996}, although the ionization states of H and He are tracked using a non-equilibrium solver. The chemical enrichment model tracks ten elements (C, O, Si, Fe, N, Ne, Mg, S, Ca, Ti), accounting for enrichment from Type II SNe, Type Ia SNe, and asymptotic giant branch (AGB) stars. Galactic outflows and wind scaling are modelled similarly as in {\sc Simba}, with different normalization factors as quoted in Table~\ref{tab:sims}. In contrast to {\sc Simba} {\sc TD} does not assume two-phase outflows, but assumes all outflows are at the ISM temperature. Photoionization feedback and reionization are treated via a self-consistent, inhomogeneous, multifrequency UVB, using a moment based radiative transfer approach on Cartesian grid~\citep{Finlator:2009}. The emissivity from galaxies is computed directly from the star formation rate of star-forming gas particles, whereas the quasar contribution depends on redshift following~\citet{Manti:2017} and on energy using the~\citet{Lusso:2015} continuum slope.

For modeling star formation, {\sc TD} adopts the sub-grid multi-phase model developed by~\citet{Springel:2003}, in which the star-forming gas (ISM) is composed of cold clouds embedded within an ambient hot medium, following~\citet{McKee:1977}. These cold clouds represent the prime repository for star formation. The mass exchange between these phases occurs through star formation, evaporation from supernovae and cloud growth due to cooling. The SFR here depends on the gas density as follows:
\begin{equation}
    \frac{d \rho_{*}}{dt}  = (1-\beta)\frac{\rho_{c}}{t_{*}}\, ,
\end{equation}
where $\rho_{c}$ is the cloud density, $t_{*}$ is characteristic time-scale to covert the $\rho_{c}$ into stellar density $\rho_{*}$, and $\beta$ is the mass fraction of stars that explodes as supernovae. Similar equation can be written for the hot medium accounting for the mass loss from clouds and supernovae energies. This model also follows the Schmidt law~\cite{Schmidt:1959} where the SFR is proportional to n$_{\rm H}^{1.5}$, with the $t_{*}$ tuned to match Kennicutt relation~\citep{Kennicutt:1998}. 
Similarly, star formation only applied for dense gas with a hydrogen number density of $n_{\rm H}$ $\geq$ 0.13 cm$^{-3}$.  {\sc TD} assumes the~\citet{Kroupa:2001} IMF, slightly different than {\sc Simba}.

The major differences between the two simulations are summarized in Table~\ref{tab:sims}.

\begin{table*}
 \caption{Major differences between {\sc Simba} and {\sc TD} }
\begin{adjustbox}{width=0.95\textwidth}
 \label{tab:sims}
 \begin{tabular}{ | l| l | l | }
  \hline
   & {\sc  \bf Simba} & {\sc \bf  TD}\\
  \hline \hline
   {\bf Star Formation} & H$_{2}$-regulated~\citep{Krumholz:2009} & Hybrid multi-phase~\citep{Springel:2003}\\[0.2cm]
   {\bf Radiative transfer} & No & Yes \\[0.2cm]
   {\bf UVB} & homogeneous, spectrum from~\citet{Haardt:2012} & inhomogeneous, multifrequency~\citep{Finlator:2009} \\[0.2cm]
   {\bf Galactic outflows}  & Yes & Yes \\[0.2cm]
   {\bf Mass loading factor $\eta(M_{*})$} &  
  \makecell[l]{ $= \begin{cases}
       9 \left(\frac{M_*}{M_{0}}\right)^{-0.317}& if M_{*} < M_{0} \\
        9 \left(\frac{M_{*}}{M_{0}}\right)^{-0.761}& if M_{*} > M_{0} 
    \end{cases}$, \\
    where $M_{0}= 5.2\times 10^{9} M_{\odot}$~\citep{Angles:2017b}. \\  This is about $\sim$ 2 $\times$ higher at $M_{*} < M_{0}$, and about $\sim$ 2 $\times$\\  lower at $M_{*} > M_{0}$ than {\sc TD}'s adopted outflows rate\\ ~\citep{Muratov:2015}. Due to the small volume ($\sim$ 22 Mpc), \\ $\eta(M_{*})$ is higher for {\sc Simba} at all relevant masses.  \\[0.2cm] }   .
    &  =  $ 2.3 \left( \frac{M_{*}}{10^{10} M_{\odot}}\right)^{-0.35}$,~\citep{Muratov:2015}.  \\[0.2cm]
    {\bf Wind velocity, $v_{w}$ } &   $ = 0.85\, v_{\rm circ}^{1.12} + \Delta v(0.25R_{\rm vir})$,~\citep[Based on][]{Muratov:2015}     &  \makecell[l]{ $ = 2.4\, v_{\rm circ}^{1.12} + \Delta v(0.25R_{\rm vir})$,~\citep[Based on][]{Muratov:2015}.  \\  This is about $\sim$ 3 times faster than {\sc Simba}'s wind velocity. \\[0.2cm]  }  \\[0.2cm]  
    {\bf Black holes } & Yes & No \\ \hline
 \end{tabular}
 \end{adjustbox}
\end{table*}

\subsection{Simulation Runs}
\begin{figure*}
\centering
\includegraphics[scale=0.5]{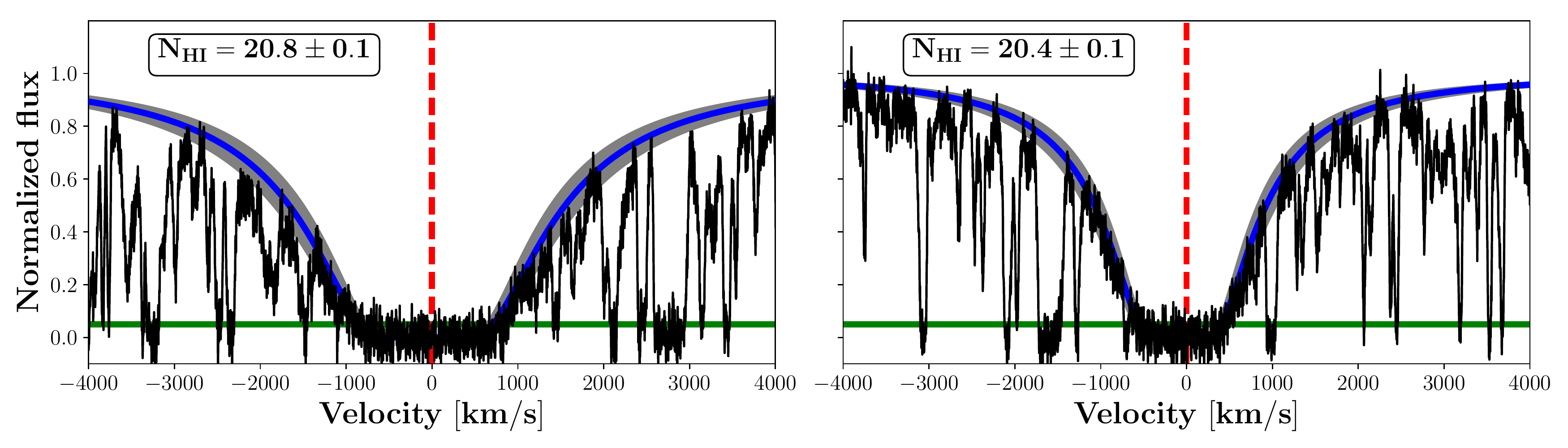}
\caption{Examples of two DLA spectra with their VP fit (solid blue) using the {\sc pyigm} package. Green and red lines represent the uncertainty and velocity centroid, respectively. The column density values are quoted in each plot.}
\label{fig:vpfits}
\end{figure*}
\begin{figure}
\centering
\includegraphics[scale=0.4]{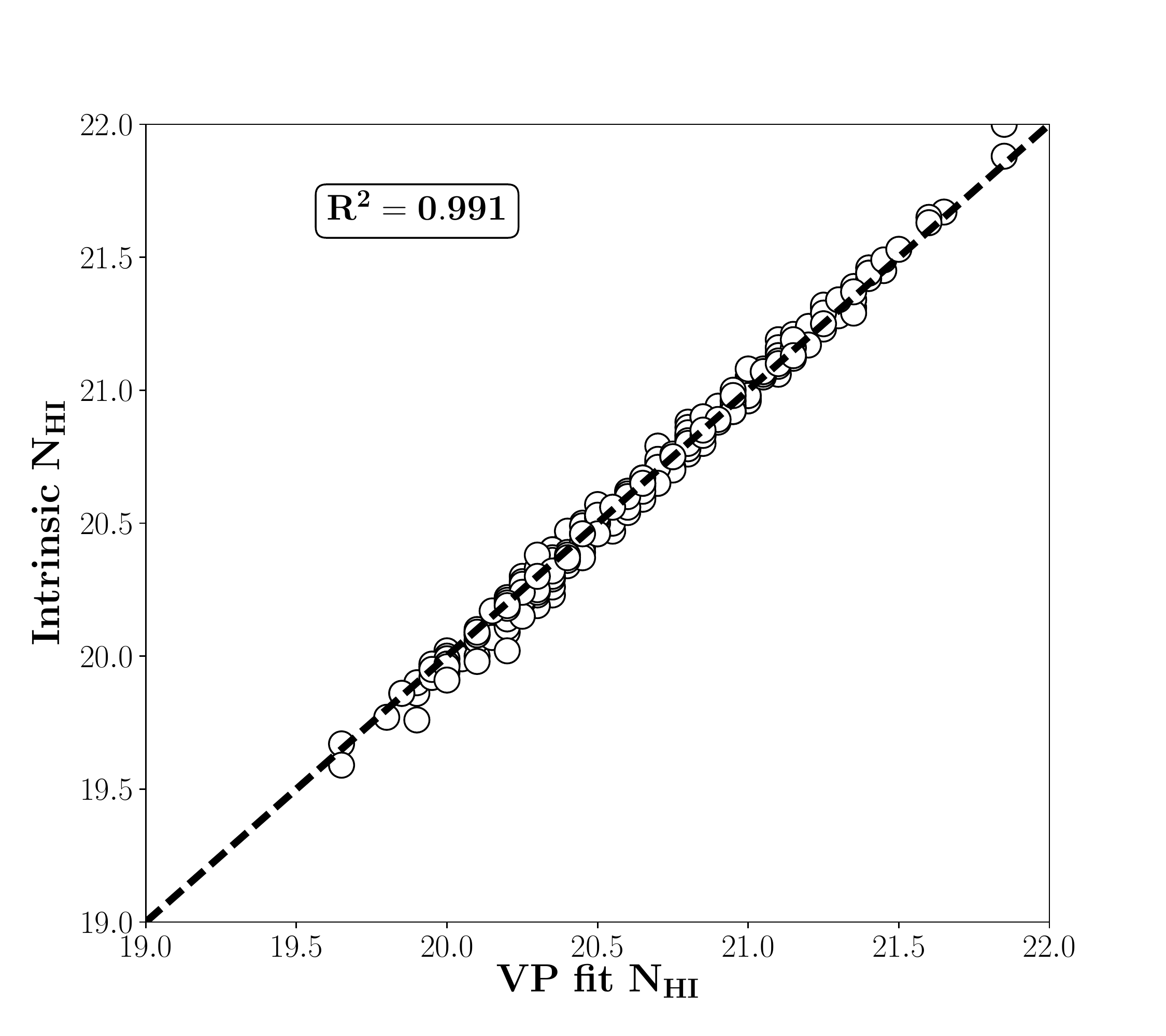}
\caption{The DLAs column density comparison at $z=3$ between values determined using a Voigt Profile (VP) fitter versus the intrinsic value from the simulation, which is computed as the total column density from pixels along the DLA profile within 500 km/s window. All DLAs are clustered along the dashed line (perfect match), which is an evident that there is no bias towards using either method, with a coefficient of determination R2 > 99\%. Hence, we proceed using the intrinsic column density value at all redshifts.}
\label{fig:NHIs}
\end{figure}
To establish a consistent comparison between these simulations, we use identical initial condition of 15 $h^{-1}$ Mpc with 2 $\times$ 640$^{3}$ dark matter and gas particles each, to run {\sc Simba} and {\sc TD} assuming the following cosmology: ($\Omega_{m}$, $\Omega_{\Lambda}$, $\Omega_{b}$, H$_{0}$,$X_{H}$) $=$ (0.3089, 0.6911, 0.0486, 67.74, 0.751), using a parallel version of N-GenIC\footnote{https://wwwmpa.mpa-garching.mpg.de/gadget/} to initialize our simulations at z=199. We then identify galaxies and halos using a friends-of-friends group finder within 3-dimensions for {\sc TD} and 6-dimensions for {\sc Simba}. In this study, we restrict our analysis to the redshift range of $z = 3 - 5$ in $dz=0.5$, where a sizeable population of DLAs have been observed, in order to test these simulations against observable DLA properties.  It is worth mentioning that our simulations' box size (15 $h^{-1}$ Mpc) is large enough to capture a representative population of DLAs although it may miss those that are associated with the most massive DLA-hosting halos. For a concrete reference point, our mass resolution is about one order of magnitude higher than the fiducial runs considered in~\citet{Bird:2014}, who modeled a 25 $h^{-1}$ Mpc volume using $512^3$ gas resolution elements. Meanwhile, our cosmological volume is about 22\% as large. Low-mass galaxies and halos and their contributions to DLAs are therefore more completely accounted for, while halos with circular velocities exceeding $\approx120$ km/s are under-represented.~\citet{Bird:2014} found that the DLA hosting halos' velocity distribution peaks at $\sim$ 100 km/s. Assuming this is correct, it suggests that our simulations may indeed be systematically deficient in DLAs associated with the most massive DLA-hosting halos. We leave exploring the relationship between the DLAs kinematics and their hosting halos/galaxies properties for a follow up work.

\subsection{The simulated DLA sample}\label{sec:dlamock}

From each simulation, we generate mock spectra following the recipe developed in~\citet{Finlator:2018}, which we review briefly here. We pass an oblique long sightline through each simulation volume, from $z$=3 to $z$=5 in intervals $\Delta z$=0.5, wrapping at the simulations boundaries, till a velocity width of 4$\times$10$^{7}$ km/s is achieved. Choosing a large velocity width is necessary to ensure detection of a representative catalog of synthetic DLAs. We next smooth the simulated density, temperature, metallicity, and velocity field onto the sightline with a pixel size of 2.67 km/s. The absorption spectra are generated using Voigt profiles. We use~\citet{Haardt:2012} photoionization background within {\sc Simba} whereas {\sc TD} uses its own inhomogeneous  photoionization background. We then compute the optical depth following~\citet{Theuns:1998}. We use an 8 km/s  full-width at half-maximum (FWHM) to smooth the generated spectra with a Gaussian filter. This results in a survey of 3 pixels per resolution element. We finally add a Gaussian noise (signal-to-noise, SNR) of 20 km/s per pixel. These values of the pixel and FWHM resolutions are similar to those values used in observations~\citep[e.g.][]{Neeleman:2013}.

We identify DLAs as \HI\ absorbers whose column density satisfies N$_{\rm \HI} \geq$  10$^{20.3}$ cm$^{-2}$. As DLAs are on the damping regime of the curve of growth, the column density relates directly to the equivalent width $W \propto \sqrt{N_{\rm \HI}}$, in which the DLA threshold translates into an equivalent width threshold of 9.17 \AA \space in the rest frame. Our procedure to identify DLAs is as follows: we first scan our long sightline of 4$\times$10$^{7}$ km/s and consider any pixels where the flux is more than $3\sigma$ below the continuum to indicate a significant absorption feature. We split the sightline into sequences of successive absorption features, assuming that each sequence represents one absorber.  We then compute the equivalent width for all these features, and regard an absorption feature as a DLA candidate if its equivalent width is equal or above the threshold. To be more conservative, we consider a smaller equivalent width value of 9.0 \AA \space as a cut-off to capture any possible DLA candidate. 

Having identified DLA candidates, we consider two methods to compute each candidate's \HI\ column density. The first is to fit a Voigt profile (VP) to its synthetic spectrum (observational method), while the second is to use the intrinsic column density as generated by the simulation (theoretical method). For the Voigt profile fit, we use the {\sc pyigm}\footnote{https://github.com/pyigm/pyigm. The package performs by-hand fit for the input DLA profile.}, a python package for the analysis of the Intergalactic Medium and the Circumgalactic Medium. Some examples for the VP fit using {\sc pyigm} are shown in Figure~\ref{fig:vpfits}. For the intrinsic column density, we compute the total \HI\ column density along the DLA profile within a window of 500 km/s about the centroid (i.e. highest column density pixel)\footnote{The choice of the window size doesn't affect the resulting column, since most of pixels along the DLA profile have a column density of 10$^{11-13}$ cm$^{-2}$, which is about 9 order or magnitude lower the threshold. The 500 km/s choice is made appropriately according to the maximum observed velocity width of DLAs.}. We now compare the outcome of the two methods on all $z=3$ simulated DLA profiles in Figure~\ref{fig:NHIs}. We notice that all DLAs are tightly clustered along the dashed line, which indicates a perfect match between the two methods, with a coefficient of determination R2 > 99\%. This shows that there isn't a bias towards using any method, and hence we proceed to use the intrinsic \HI\ column density method at all redshifts. 
\section{Simulation comparison}\label{sec:simcom}
We begin by comparing {\sc Simba} and {\sc TD} in terms of their dense gas distribution, star formation, and UVB treatment, in order to set the stage for understanding the physics driving DLA properties.

\subsection{ISM Density Distribution}
\begin{figure}
    \centering
    \includegraphics[scale=0.4]{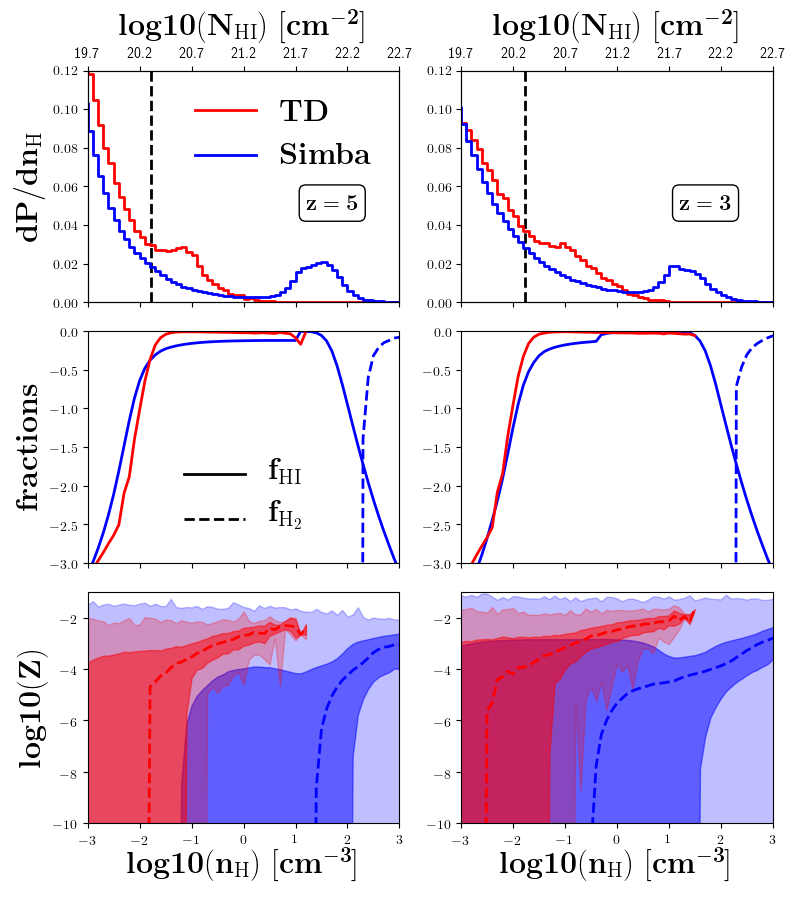}
    \caption{Density distribution in {\sc Simba} (blue) and {\sc TD} (red)  at $z=5$ (left) and $z=3$ (right). Top axes are labeled with \HI\ column density computed from H number density using Equation~\ref{eq:NHI_HI}. Top: hydjrogen number density distribution. {\sc TD} has more gas at $0.01<n_{\rm H}<1\,  $cm$^{-3}$ than {\sc Simba} where the majority of DLAs are expected to form. This is shown by the dashed vertical line, corresponding to the DLA column density threshold. The density distribution in {\sc Simba} extends to much higher densities $\sim$ 10$^{3}$ cm$^{-3}$ that are two orders of magnitude higher than the highest in {\sc TD},  due to the lower wind speed and H$_{2}$-regulated SF model. Middle: the 50$^{th}$ percentile of neutral and molecular hydrogen fractions as a function of hydrogen number density.  Both simulations show  a sharp increase in f$_{\rm \HI}$ (solid lines) at 0.01 cm$^{-3}$. The f$_{\rm \HI}$ in {\sc Simba} decreases rapidly at 10 cm$^{-3}$ where most of the gas is being converted to molecular as indicated by the increase of f$_{\rm H_{2}}$ (dashed line). {\sc TD} doesn't form molecular hydrogen.  Bottom: total metallicity as a function of hydrogen number density. Dark shaded areas encompass the 15.9$\%$ and 84.1$\%$ (i.e. $\sim$ 1-$\sigma$ level) of the metallicity distribution whereas the dashed lines represent the 50$^{th}$ percentile. The faster winds in {\sc TD} eject metals to larger distances and less denser regions than in {\sc Simba}, indicating {\sc TD}'s ability to form more metal rich DLAs. }
    \label{fig:gas_dist}
\end{figure}
Figure~\ref{fig:gas_dist} shows gas density distribution in {\sc Simba} and {\sc TD} at the highest and lowest redshifts considered in this study ($z=5,3$). In the top panel, we compare the hydrogen number density n$_{\rm H}$  distribution in the two simulations. While both simulations have started from the same initial condition, we see a big difference in the n$_{\rm H}$ distribution, particularly at the dense end (i.e. the shelf-shielding regime). In the top axis, we show a rough estimate of the \HI\ column density corresponding to the hydrogen number density, following~\citet{Schaye:2001a} as:
\begin{equation}\label{eq:NHI_HI}
    N_{\rm \HI} \sim 1.6 \times 10^{21} {\rm cm}^{-2}\, n_{\rm H}^{1/2}\, T_{4}^{1/2}\, \left(\frac{f_{g}}{0.17}\right)^{1/2} , 
\end{equation}
where $f_{g}$ is the universal baryon fractions of absorbers $\frac{\Omega_{b}}{\Omega_{m} }$, and the temperature $ T_{4}  \equiv T/10^{4}$ is set to unity, ignoring collision ionization. This equation provides an approximate estimate of the \HI\ column density at high densities and high neutral fraction values. The vertical dashed lines shows the DLA column density threshold ($N_{\rm \HI}  = 2\times 10^{20}$ cm$^{-2}$). There is clearly more gas in {\sc TD} than {\sc Simba} above the DLA threshold and below  $N_{\rm \HI} \sim 10^{21} {\rm cm}^{-2}$, in the hydrogen number density range  $0.01 - 1$ cm$^{-3}$, where the majority of DLA population exists. This might be due to the lower outflow rate (about $\sim 2 \times$ less than {\sc Simba}; see Table~\ref{tab:sims}) that {\sc TD} applies at the intermediate stellar masses as well as the {\sc TD}'s faster winds (about $\sim 3 \times$ higher than {\sc Simba}).  The density distribution in {\sc Simba} extends to 10$^{3}$ cm$^{-3}$, which is roughly two orders of magnitude higher than the highest gas density in {\sc TD}. This could reflect either {\sc Simba}'s low outflows speed (about $3 \times$ less than {\sc TD}; see Table~\ref{tab:sims}). Alternatively, it could reflect {\sc Simba}'s tendency to suppress star formation in low-metallicity gas due to the H$_{2}$-regulated star formation model.  

In the middle panel, we show the 50$^{th}$ percentile of neutral f$_{\rm \HI}$ and  molecular  f$_{\rm H_{2}}$ fractions  as a function of hydrogen number density by the solid and dashed lines respectively. Both simulations indicate a sharp increase in f$_{\rm \HI}$ at gas densities near 0.01 cm$^{-3}$, which is due to self-shielding ~\citep{Rahmati:2013}. At densities higher than 10 cm$^{-3}$, the neutral fraction f$_{\rm \HI}$ in {\sc Simba} drops suddenly, marking the transition to molecular gas as shown by the rapid increase of  f$_{\rm H_{2}}$. This is a distinct feature of {\sc Simba}, which is a consequence of implementing the H$_{2}$-regulated recipe to form stars. This feature is less important in {\sc TD}, where the hybrid multi-phase model triggers star formation based solely on the total gas density. The total metallicity as a function of gas density in our simulations is depicted in the bottom panel, in which the effect of wind speed is prominent. Dark shaded areas encompass the 1-$\sigma$ level about the 50$^{th}$ percentile that is represented by the dashed lines. The faster winds in {\sc TD} eject metals to larger distances and less denser regions than in {\sc Simba}. This indicates that {\sc TD} would be able to form more metal rich DLAs than {\sc Simba}.

\subsection{Star Formation and Stellar Mass Growth}

\begin{figure}
    \centering
    \includegraphics[scale=0.3]{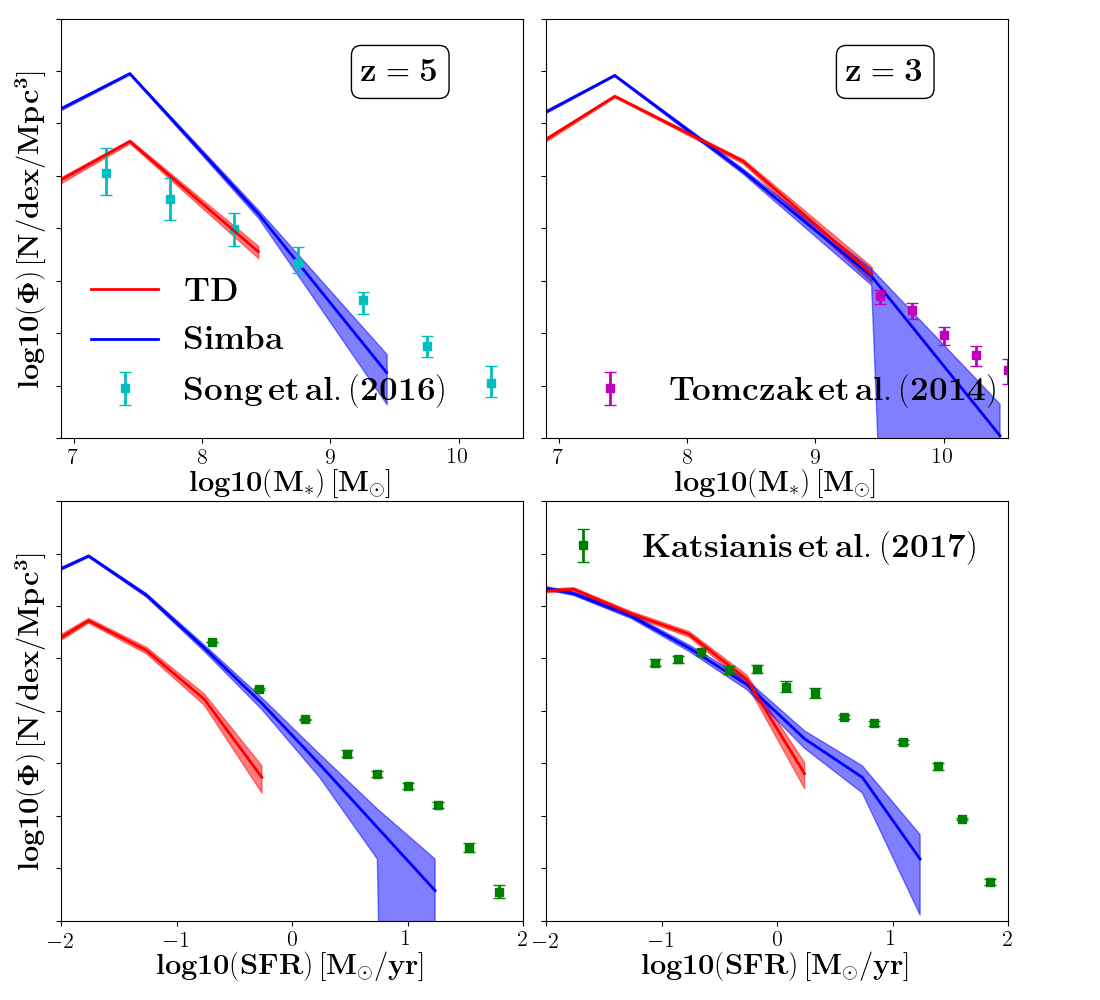}
    \caption{Comparison between {\sc Simba} (blue) and {\sc TD} (red) against the measured stellar mass functions (top) and SFR functions (bottom) at $z=5$ (left) and $z=3$ (right). Shaded areas reflect the Poisson errors. At $z=3$, both simulations, regardless of the adopted star formation model and feedback effects, have relatively similar stellar mass and SFR functions at M$_{*} \leq 10^{9.5} M_{\odot}$ and  SFR $\leq$ 1 $M_{\odot}/yr$. The stellar mass function in {\sc Simba} extends to larger masses up to $\sim 10^{10.5} M_{\odot}$ that are one order of magnitude larger than the highest stellar mass resolved in {\sc TD}, owing to the low wind speed in {\sc Simba} and the H$_{2}$-regulated SF model. At higher redshift $z=5$, {\sc Simba} produces stellar mass and SFR functions that are $\sim \times$ 3 higher than those of {\sc TD}, and the ability of {\sc Simba} to resolve larger stellar systems is still seen in this epoch. This indicates that {\sc TD} applies stronger feedback ($\sim3\times$ faster wind speed than {\sc Simba}) that suppresses SFR at small stellar masses as compared with {\sc Simba}. The agreement between our simulations at $z=3$ is largely due to {\sc TD}'s weaker UVB effects as depicted in Figure~\ref{fig:flux}.   In general, our simulations both are within $\sim$ 1-3 $\sigma$ level of measurements.}
    \label{fig:SMF}
\end{figure}

Before examining the DLA population, we would like to ensure that both TD and {\sc Simba} are producing a reasonable population of high-$z$
galaxies.  To assess the galaxy population, we compare their stellar mass and star formation rate functions at $z=5$ and $z=3$ to each other as well as to observations. Shaded areas reflect the Poisson errors.  Differences in stellar growth are most sensitive to variations in feedback prescriptions. Figure~\ref{fig:SMF} shows the stellar mass (top panels) and SFR (bottom) functions in our simulations at $z=5$ and $z=3$. At higher redshifts, {\sc Simba} forms more stars than {\sc TD}, resulting in higher stellar mass and SFR functions by a factor of $\sim 3$.  Even though the outflow mass loading factor is $\sim\times 2$ higher in {\sc Simba} than TD which should reduce stellar growth, the wind velocities are $\sim\times 3$ lower, which evidently yields significantly more wind recycling at early times that counters this and enables rapid early growth.

\begin{figure}
    \centering
    \includegraphics[scale=0.6]{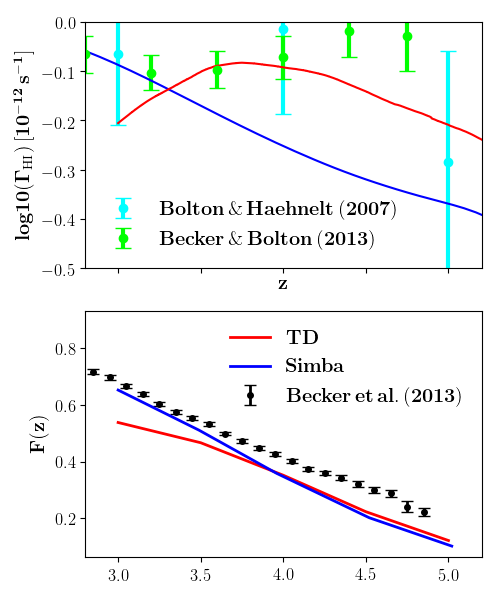}
    \caption{Top: \HI photoionization rate evolution over redshift. {\sc TD} models a spatially inhomogeneous UVB, whereas {\sc Simba} adopts the homogeneous~\citet{Haardt:2012} UVB. At $z>3.5$, {\sc TD} has about $\sim 1.5 \times$ higher photoionization rate than {\sc Simba}, and yet the mean transmitted flux is the same in both simulations. 
    This might also be due to the {\sc TD}'s high photon escape fraction at these epochs. Bottom: Mean transmitted flux in Ly-$\alpha$ forest comparison between {\sc Simba} (blue), {\sc TD} (red) and the measurements by~\citet{Becker:2013}. Both simulation have lower transmitted flux than observed, indicating that the UVB is weak, temperature is low, and the opacity is high. {\sc TD} systematically under-produces the~\citet{Haardt:2012} predicted mean transmission for $z$<4, which partially contributes to the DLA over abundance at these epochs.}
    \label{fig:flux}
\end{figure}

By $z=3$, both models are in fairly good agreement.  The main difference is that, owing to its rapid early growth, {\sc Simba} produces some fairly large galaxies with $M*\ga 10^{10}M_\odot$ with SFR$\ga 10 M_\odot$yr$^{-1}$, whereas TD does not.  As such, the cold gas content that gives rise to DLAs in these two models can be robustly compared, as there has not been a strong difference in the amount of cold gas converted into stars. On the other hand,~\citet{Pawlik:2009} (and later confirmed by ~\citealt{Finlator:2011}) have previously shown that outflows and the UVB couple nonlinearly to suppress inflows into halos at all masses, which in turn suppresses star formation. Put differently, the UVB's impact on galaxy growth is stronger in the presence of outflows. It is reasonable to suppose that, if the UVB thus suppresses inflows, then it also suppresses the DLA abundance; we will return to this point below. At this epoch $z=3$, {\sc TD} is able to form more stars, similar to {\sc Simba}, due to {\sc TD}'s weaker UVB (see Figure~\ref{fig:flux}).

Comparing the models to observations, there are clearly some discrepancies, and neither model agrees well with all the data.  The most reliable observations among those shown here are the \citet{Tomczak:2014} stellar mass functions, as they are obtained from rest-optical observations.  The \citet{Song:2016} results are derived from rest-UV data, from which getting a stellar mass can be sensitive to many uncertainties regarding stellar populations and IMF.  Similarly, the \citet{Katsianis:2017} results come from rest-UV data which are sensitive to extinction in large galaxies, and stellar populations in small galaxies.  In general, both models fall short in predicting the abundance of the highest SFR and $M_*$ galaxies.  This is in part explainable by the small volume; in the case of \simba, the $100\hmpc$ run presented in \citet{Dave:2019} agrees well with the stellar mass functions out to $z=6$.  However, it is unclear that simple cosmic variance explains all the discrepancy.  Given the various uncertainties in both the observations and the simulations, we will not draw strong conclusions regarding any discrepancies, but rather defer a more careful comparison of the galaxy population in observational space to future work.

\subsection{Ultraviolet Ionizing Background (UVB) treatment}

As described earlier, {\sc TD} self-consistently implements a radiative transfer to model a multi-frequency spatially-inhomogeneous UVB, while {\sc Simba} includes no radiative transfer routine to generate its own UVB field but rather uses the homogeneous~\citet{Haardt:2012} background. The UVB spatial distribution and strength affects the neutral fraction along a sightline and hence the Ly-$\alpha$ transmitted flux. We explore the differences in the UVB between these two simulations by examining the \HI\ photo-ionisation rate $\ghi(z)$, and its resulting impact on the mean transmitted flux in the \lya\ forest.  While this does not significantly impact DLAs since they come from dense self-shielded gas, it provides an interesting comparison between the self-consistently generated $\ghi$ in TD versus that in \citet{Haardt:2012}.

Figure~\ref{fig:flux}, top panel, shows the \HI\ photoionization rate $\ghi$ as a function of redshift in {\sc Simba} (blue) and TD (red) against inferred  measurements from the Ly-$\alpha$ forest by ~\citet{Bolton:2007} and~\citet{Becker:2013b}, shown in cyan and green respectively. 

Both simulations have a $\ghi$ that is fairly consistent with observations, given the large uncertainties at high redshifts. {\sc Simba} $\ghi$ taken from~\citet{Haardt:2012} shows a steady increase from $z=5\to 3$.  The self-consistent modelling of RT in {\sc TD} predicts a higher $\ghi$ by a factor of $\sim\times 1.5$ versus {\sc Simba} at $z\ga 4$. This might be due to {\sc TD}'s high photon escape fraction at these epochs, compared to what is assumed in \citet{Haardt:2012}.  At $z\la 3.5$, TD predicts that $\ghi$ turns over.  This may be because of a lack of high mass galaxies in the small volume to self-consistently generate ionising photons, or else the small contribution from AGN which begin to be an important contributor to $\ghi$ at these epochs.  

Figure~\ref{fig:flux}, bottom panel, shows the mean transmitted flux in the Ly-$\alpha$ forest {\sc TD} and {\sc Simba} at these redshifts as the red and blue lines, respectively, versus observations as compiled by~\citet{Becker:2013}.  Following~\citet{Becker:2013}, we define the IGM as pixels with column density N$_{\rm \HI} < 10^{19}$cm$^{-2}$, to compute the mean transmitted flux from only diffuse and high-ionized absorbers\footnote{We have checked that including the DLA profiles doesn't change the resulting mean transmitted flux, since DLAs are rare.}.  Broadly, the mean transmitted flux is most sensitive to marginally saturated lines, i.e. $\nhi\sim 10^{14}\cdunits$, since above this column density the lines enter the logarithmic portion of the curve of growth.

The mean transmitted flux increases with time, as the Universe expands and its density drops. The rate of increase is similar in both models, and is comparable to observations.  However, we see that both simulations somewhat under-produce the mean Ly-$\alpha$ transmission at all redshifts, which suggests the UVB in both simulation is slightly too weak. It has been noted previously that the~\citet{Haardt:2012} UVB under-produces mean transmission~\citep{Finlator:2018,Bosman:2018}. \citet{Gnedin:2017} find qualitatively similar results within a different radiative hydrodynamic simulation.  The top panel suggests that the predicted UVB in TD is consistent with observations, so it is not entirely clear why the mean transmitted flux is different.  For {\sc Simba}, the low photo-ionisation rate directly translates to too little transmission by a similar factor.

At $z<4$, the mean transmitted flux starts to increase in {\sc Simba} and becomes in a good agreement with measurements at $z=3$. Meanwhile, by $z=3$, {\sc TD} under-predicts the observed mean transmission by a factor of $\sim$ 1.5, indicating that the UVB is too weak and opacity is too high.  This may play into the DLA statistics at some level.

\section{DLA Abundance}\label{sec:dlaabund}
\begin{figure*}
    \centering
    \includegraphics[scale=0.35]{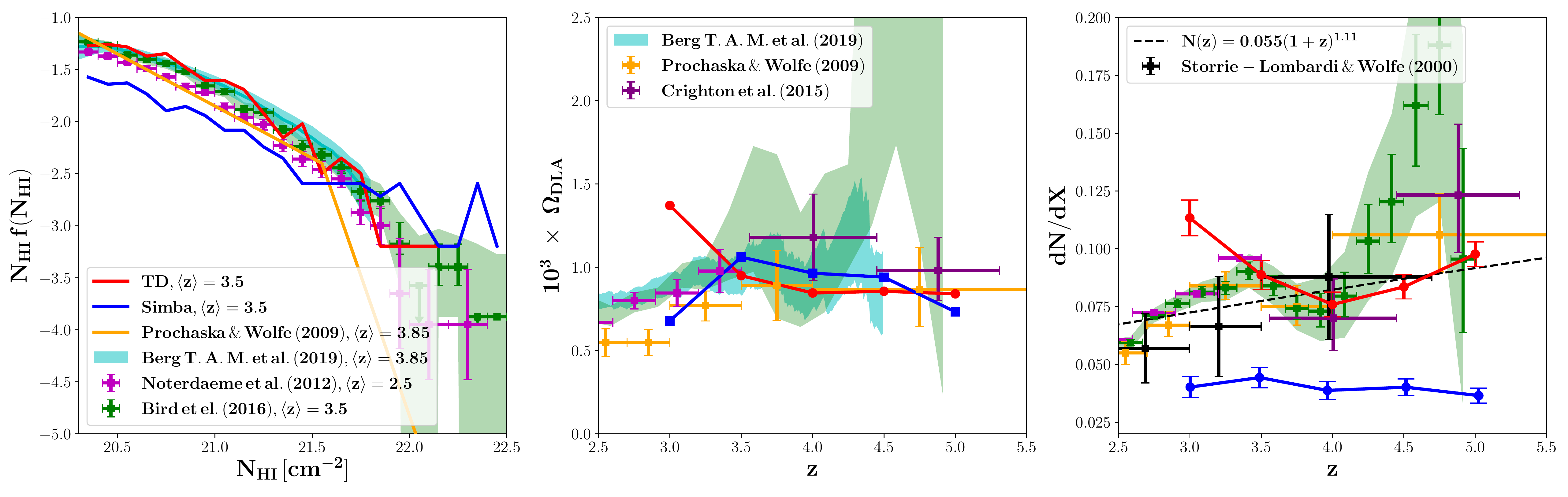}
    \caption{Comparison between {\sc Simba} and {\sc TD} simulations with measurements by~\citet[cyan]{Berg:2019},~\citet[green]{Bird:2017},~\citet[magenta]{Noterdaeme:2012},~\citet[orange]{Prochaska:2009},~\citet[purple]{Crighton:2015} and~\citet[black]{Storrie-Lombardi:2000}, in terms of the column density function distribution (left), the \HI\ density evolution (middle), and the abundance evolution of DLAs (right). Error bars and shaded areas represent the 1-$\sigma$ and 2-$\sigma$ levels of measurements, respectively. Mean redshift of each sample is quoted in the legend.  {\sc Simba} under-predicts the observed abundance by a factor of 2, although uncertainty increases with redshift. {\sc TD} produces a consistent abundance with observations, albeit DLAs are over-produced at $z=3$ by a factor of 2. Similar trends are seen in the column density distribution function, except {\sc Simba}  resolves higher column density systems at $\geq$ 10$^{22}$ cm$^{-2}$, which drives the increase seen for the DLAs \HI\ density at the mean redshift. DLAs under-production are partially due to the difference in the outflows feedback strength and perhaps the star formation recipe.}
    \label{fig:all}
\end{figure*}

In this section, we test the viability of our simulations to reproduce the DLA observations for abundance evolution, the column density distribution, and neutral density evolution.  
\subsection{DLA Abundance evolution}
The DLA abundance ($\frac{dN}{dX}$) is the number of DLAs identified in the simulation volume at redshift $z$ for a survey width $\Delta v$ corresponding to an absorption length $dX$, alternatively called, the line density of DLAs per comoving absorption length $dX$. The DLA abundance is mainly driven by the abundance of the low column density systems, since the high column density systems are quite rare.

To compute the $dX$ for our mock survey at a redshift $z$, we first convert the survey velocity width ($\Delta v = 4 \times 10^{7}$ km/s) into the redshift interval using the relation: $dz = \Delta v (1+z)/c$, where c is the speed of light. The absorption length $dX$ is then computed as follows:
\begin{equation}
\frac{dX}{dz}  =  (1+z)^{2} \frac{H_{0}}{H(z)} = (1+z)^{2} / \sqrt{\Omega_{m} (1+z)^{3} + \Omega_{\Lambda}}.
\end{equation}

We now compare the $\frac{dN}{dX}$ from {\sc Simba} (blue errorbars) and {\sc TD} (red errorbars) with observations in the right panel of Figure~\ref{fig:all}.  The black errorbars and their dashed black line fit are the early measurements by \citet{Storrie-Lombardi:2000}, orange errorbars are complied by~\citet{Prochaska:2009} using the SDSS DR5,  magenta errorbars are measurements by~\citet{Noterdaeme:2012},  green errorbars are the measurements by~\citet{Bird:2017} using SDSS DR12 survey~\citep{Garnett:2017}, purple errors are measurements by~\citet{Crighton:2015} using the Giant Gemini GMOS survey, and cyan errors are the most resent measurements reported by~\citet{Berg:2019} using $XQ-100$ survey. We find that {\sc Simba} under-predicts the observed DLA abundance at $z$>3 approximately by a factor of 2, but still within 1$-$3 $\sigma$ level of the observations, particularly at $z>4$ where the observational uncertainty is large. The TD simulation, on the other hand, is more consistent with the current DLA abundance estimates at $z \geq$ 3.5. Below this redshift, the TD simulation starts to over-produce DLAs as seen at $z$=3. This over-production of DLAs might correlate with the weak UVB and the high IGM opacity that {\sc TD} predicts at $z$=3, as seen in Figure~\ref{fig:flux}. This effect has previously been found in~\citet{Bird:2014}, where they conclude that an increased UVB amplitude reduces the DLA cross section, and suppresses the DLA abundance. 

Although these simulations have the same mean transmitted flux at $z \geq 4$, still {\sc Simba} under-produces DLAs by a factor of 2, as compared with {\sc TD} and observations, which implies that the UVB treatment cannot solely explain the differences seen in DLA abundance evolution. However, the remarkable differences seen in the ISM density distribution in Figure~\ref{fig:gas_dist} as well as the stellar mass functions in Figure~\ref{fig:SMF} at $z=5$, all together indicate that the implemented star formation recipes and feedback effects mainly contribute to the under/over production of DLAs as seen in Figure~\ref{fig:all}.  This can be explained by the difference in the outflows strength. Comparing with {\sc Simba}, it appears that reducing the outflows rate by a factor of $2$ as well as boosting the wind speed by a factor or $3$ (see Table~\ref{tab:sims}) both suppress the SFR at $z=5$, and then induces more DLAs in {\sc TD}. This effect has been previously noted in~\citet{Faucher:2015}, where stronger outflows feedback were found to suppress SFR, and enhance the \HI\ covering fractions.

\subsection{DLA Column density distribution}
The column density distribution function (CDDF) is defined as the number of DLAs per unit column density ($dN_{\rm \HI}$) per unit comoving absorption length ($dX$). We compare the CDDF from {\sc Simba} and {\sc TD} simulations with measurements by~\citet[cyan]{Berg:2019}, ~\citet[green]{Bird:2017}, ~\citet[magenta]{Noterdaeme:2012},~\citet[purple]{Crighton:2015}, and~\citet[orange]{Prochaska:2009} in the left panel of Figure~\ref{fig:all}.
For consistent comparison, we only consider DLAs from simulations at $z=3,3.5,4.0$, with $\langle z \rangle =3.5$ that is nearly equal to the mean redshift of these various measurements as quoted in the legend.

Here we see similar trends as with the $\frac{dN}{dX}$ panel. At $N_{\rm \HI} < 10^{22}$cm$^{-2}$, {\sc TD} produces a consistent CDDF with observations, whereas {\sc Simba} is lower by a factor of 2. This result was anticipated by Figure~\ref{fig:gas_dist}, where stronger outflows (as in {\sc TD}) boost the hydrogen number density PDF at DLA column densities. In contrast, suppressed outflows (as in {\sc Simba}) leave more gas in high column density systems ($ > 10^{22}$cm$^{-2}$). This increase at high column densities is too small to affect the overall abundance ($\frac{dN}{dX}$), which is dominated by low-column systems. Nonetheless, it boosts the CDDF appreciably at higher columns.

Our simulations as well as some observations ~\citep{Berg:2019,Bird:2017} show no turn over for the CDDF at high column density end at about $N_{\rm \HI} = 10^{21.5}$cm$^{-2}$, which was initially suggested by~\citet{Schaye:2001}, and later predicted by~\citet{Altay:2011} and \citet{Bird:2014}, to occur due to molecular hydrogen transition that is responsible to set the maximum \HI\ column density, and hence steepening the CDDF. This turn over was also previously measured, for example, by~\citet{Prochaska:2009}, in which a double power law was used to fit the CDDF as shown by the orange solid line. Unlike {\sc TD}, {\sc Simba} explicitly includes molecular formation to model star formation via the $H_{2}$-regulated SFR recipe (see Figure~\ref{fig:gas_dist}), and yet show no turn over at the high column density. It remains interesting to test whether the observed turn over at high column densities appears in larger simulation volumes that capture more massive halos and more high column density DLAs, whose observational selection can be affected by a dust bias~\citep{Krogager:2019}.

\begin{figure}
    \centering
    \includegraphics[scale=0.5]{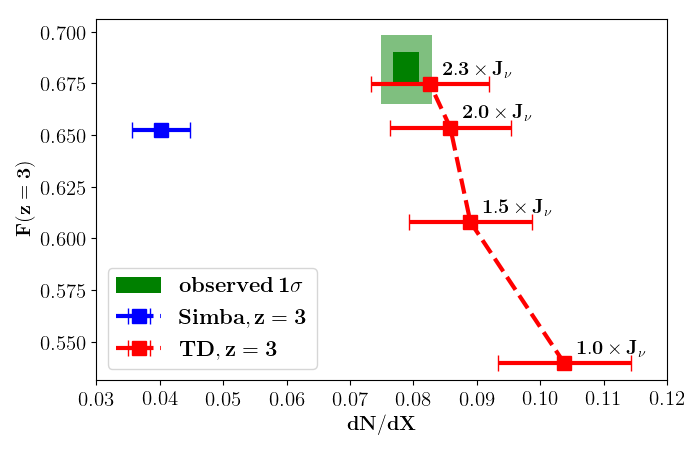}
    \caption{Mean transmitted flux in Ly-$\alpha$ forest from {\sc TD} (red square) as a function of DLA abundance at $z=3$, for different UVB scale factors as quoted next to each point. The output from {\sc Simba} at $z=3$, without scaling the~\citet{Haardt:2012} UVB, is shown by the blue square. Error bars reflect Poisson uncertainty. The dark and light green shaded box correspond to 1-$\sigma$ and 2-$\sigma$ uncertainties from~\citet{Becker:2013} on the y-axis and ~\citet{Bird:2017} on the x-axis, respectively.  The UVB is re-scaled in post-processing assuming ionization equilibrium. A stronger UVB boosts the Ly-$\alpha$ transmitted flux and reduces the DLA abundance.  {\sc TD} reproduces both observations at z=3 when we boost the predicted UVB amplitude by a factor of $2.3\times$. In contrast, scaling the UVB in {\sc Simba} would only improve the model prediction for either the Ly-$\alpha$ transmitted flux or DLA abundance. Stronger outflows feedback is clearly needed in the case of {\sc Simba} in order to approach the green box. }
    \label{fig:uvb_effs}
\end{figure}

\begin{figure*}
    \centering
    \includegraphics[scale=0.44]{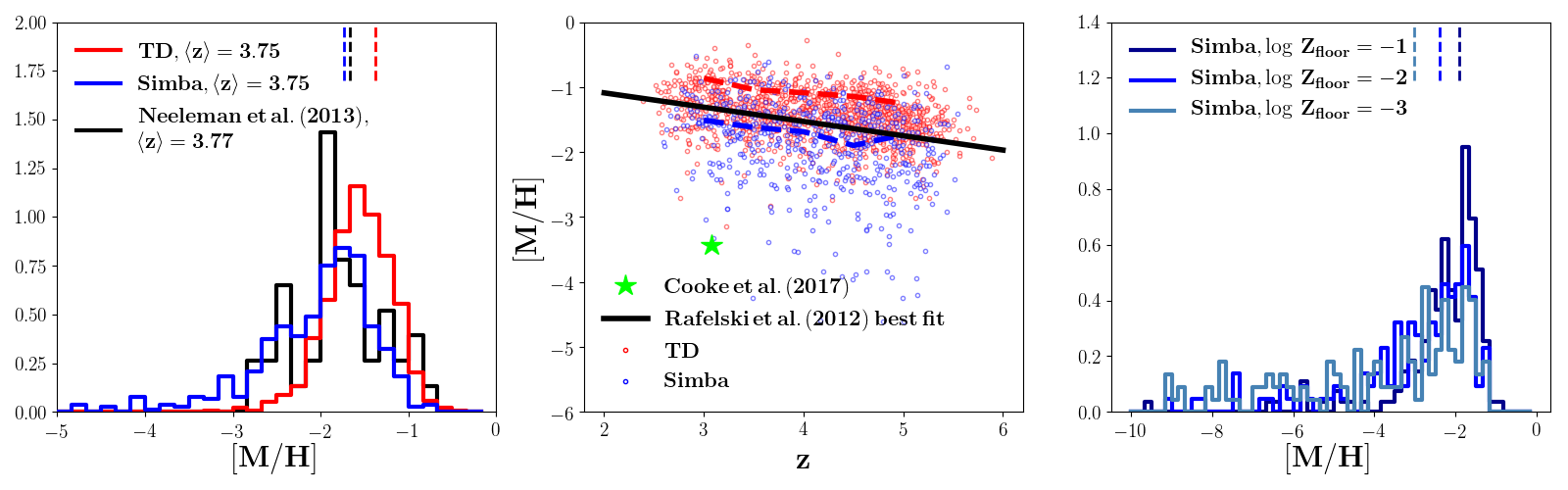}
    \caption{Comparison between the predicted and the observed metallicity distribution and evolution. Left: the metallicity distribution from {\sc Simba} (blue) and {\sc TD} (red) as compared with~\citet{Neeleman:2013} data. Middle: the metallciity evolution over redshift from our simulations compared with~\citet{Rafelski:2012}. Right: the effect of chaning $Z_{\rm floor}$ in the H$_{2}$-regulated SFR model in {\sc Simba} on the metallicity distributions. Both simulations agree with the observed metallicity evolution, predicting a negative correlation between metallicity and redshift. {\sc Simba} predicts the existence of extremely low metallcitiy DLAs that partially disappears at high $Z_{\rm floor}$ values. {\sc Simba} provides an opportunity to study the nature of the low metallicity DLA systems, whereas {\sc TD} enables studying the nature of dusty DLAs at high columns. }
    \label{fig:MH}
\end{figure*}
\subsection{Neutral density evolution in DLAs}
We define total \HI\ density in DLAs, $\Omega_{\rm DLA}$ as follows:
\begin{equation}
  \Omega_{\rm DLA} = \frac{m_{p}\,H_{0}}{c\, \rho_{c}}  \frac{ \sum_{i} N_{\rm \HI,i}}{dX}\,  ,
\end{equation}
where $m_{p}$ is the proton mass and $\rho_{c}$ is the critical density. Unlike the $\frac{dN}{dX}$, the $\Omega_{\rm DLA}$ is weighted towards high column density systems. We now compare the $\Omega_{\rm DLA}$ evolution over redshift from our simulations with measurements by~\citet[orange]{Prochaska:2009},~\citet[magenta]{Noterdaeme:2012},~\citet[purple]{Crighton:2015},~\citet[green]{Bird:2017}, and~\citet[cyan]{Berg:2019} in the middle panel of Figure~\ref{fig:all}. We here find that both simulations are consistent with measurements, except {\sc TD} over-predicts the \HI\ density in DLAs at $z$=3, which is consistent with over-production of DLAs at this redshift (see the right panel of Figure~\ref{fig:all}). While it under-estimates the CDDF and $\frac{dN}{dX}$ by a factor of $\sim$ 2, {\sc Simba} is in a good agreement with $\Omega_{\rm DLA}$ observations due to the simulation ability to resolve more high column density systems as seen in the CDDF panel. Note that if we use the same column density cut-off when computing $\Omega_{\rm DLA}$ for {\sc TD} and {\sc Simba}, particularly if we use the maximum column density obtained in {\sc TD} as a cut-off for both simulations, {\sc Simba} then under-predicts the $\Omega_{\rm DLA}$ by a factor or 2 at these redshifts. The fact that both simulations have relatively similar \HI\ density evolution over redshift and yet {\sc Simba} under-produces DLAs is suggestive that DLA cross section in {\sc Simba} is lower than in {\sc TD}. We leave exploring the DLAs connection to their hosting halos/galaxies properties to a follow-up work.

\subsection{Effect of UVB on Ly$\alpha$ transmitted flux and DLA abundance}

In order to determine whether the overproduction of DLAs in {\sc TD} at $z=3$ can be attributed entirely to the weak UVB, we re-scaled the simulated UVB in post-processing and re-generated our synthetic DLA catalog under the assumption that the gas is in ionization equilibrium with the adjusted UVB. The gas temperature was left unchanged, and the effect of self-shielding was re-computed for consistency with the new UVB. Figure~\ref{fig:uvb_effs} summarizes the results of this experiment. This figure shows the mean transmitted flux in Ly-$\alpha$ forest as function of DLA abundance for {\sc TD} (red squares) for different UVB scale factors as quoted next to each point. {\sc Simba}'s prediction at $z=3$, without scaling the~\citet{Haardt:2012} UVB, is shown by the blue square.  Error bars reflect the Poission uncertainty. Dark and light green shaded boxes show the corresponding 1-$\sigma$ and 2-$\sigma$ of observations by~\citet{Becker:2013} and~\citet{Bird:2017}, respectively. 

In qualitative agreement with~\citet{Bird:2014}, we find that scaling our simulated UVB amplitude up by a factor of 2.3 brings both {\sc TD}'s predicted mean transmission in the Ly$\alpha$ forest and DLA abundance into agreement with observations. In contrast, while adopting a weaker UVB than the~\citet{Haardt:2012} model might alleviate the discrepancy between {\sc Simba} and DLA abundance observations, it would exacerbate the discrepancy between the {\sc Simba} predictions and the Ly$\alpha$ transmitted flux measurements. This indicates that stronger outflows feedback is needed for {\sc Simba} to reproduce both measurements.  These comparisons support previous suggestions that the DLA abundance is sensitive both to the efficiency with which galaxies eject gas into their CGM, and to the UVB amplitude.

\section{DLA Metallicity}\label{sec:dlamet}

In this section, we use the observed DLA metallicity as a probe to the star formation models implemented in our simulations. We define the DLA metallicity [M/H] as the N$_{\rm \HI}$-weighted metallicity within a window of 500 km/s about the DLA centroid. To compute the metallicity in pixels along the sightline, we consider all metal species (C,N,O,Ne,Mg,Si,S,Ca,Fe) in {\sc Simba} and conisder only 4 metals (C, O, Si, Mg) in {\sc TD} that are normalized differently to solar by their corresponding fractions of 0.0134 and 0.00958 respectively. However, these 4 metals already represent more than 70\% out of total metallicity tracked in {\sc TD}, and it has been shown that with these 4 metals, {\sc TD} reproduces the DLA metallicity measurements at z$\sim$ 5~\citep{Finlator:2018}. 

We compare the predicted and observed DLA metallicity distribution and evolution over redshift in Figure~\ref{fig:MH}. We begin our discussion here with the metallicity evolution over redshift as seen in the middle panel, where DLAs from {\sc TD} and {\sc Simba} are shown by red and blue circles, respectively, and the dashed lines show their corresponding N$_{\rm \HI}$-weighted mean metallicity out of all DLAs at each redshift, following~\citet{Rafelski:2012}. Only for display purpose, we have added a scatter drawn from gaussian distribution of zero mean and 0.25 standard deviation around each DLA redshift.  Best fit line from measurements  by~\citet{Rafelski:2012} is represented by the black solid line which indicates that the DLA metallicity decreases with increasing redshift. The most-poor DLA metallicity reported by~\citet{Cooke:2017}  at z$\sim$ 3 is shown by the green lime star.  We here see that both simulations agree with the measured evolution and their corresponding mean metallicity (solid) lines have a similar negative slope to the measurements. {\sc TD} has a higher running median amplitude than {\sc Simba} by a factor of 1.3. {\sc Simba} predicts the existence of DLAs with extremely low-metallicities, that are much lower than that of the most metal-poor DLA observed to-date~\citep{Cooke:2017}, as shown by blue circles below the green lime star. This also clearly appears in the metallicity PDF (left and right panels), indicating that {\sc Simba} provides an opportunity to study the nature of the extremely low metallicity DLA systems such as that of~\citet{Cooke:2017}. The formation of these metal-poor DLAs in {\sc Simba} might be due to the weaker winds that induce a rapid drop of metallicities at higher densities as seen in Figure~\ref{fig:gas_dist}.

In the left panel, we show the metallicity distribution from {\sc Simba} (blue) and {\sc TD} (red) as compared with~\citet{Neeleman:2013} data. To establish a proper comparison between our simulations and observations, it is important here to match the mean redshift between samples, due to the metallicity evolution with redshift \citep{Prochaska:2003}. We exclude DLAs with z$\leq$3 and the only DLA at z$\sim$5 from~\citet{Neeleman:2013} observational sample, resulting in mean redshift of $\langle z \rangle = 3.77$. From our simulations we exclude $z$=5 DLAs to obtain a mean redshift of $\langle z \rangle = 3.75$. The minimum and maximum metallicity  in~\citet{Neeleman:2013} sample here are [M/H] $= -2.56$ and $-0.64$ respectively.  We see similar trends here that {\sc Simba} has more low-metallicity DLAs than {\sc TD} and the measurements. Specifically, {\sc Simba} and {\sc TD} have about $\sim 13\%$ (58) and $0.3\%$ (4) out of all DLAs that have metallicities lower than the observed minimum.  Interestingly, {\sc TD} over-produces and {\sc Simba} under-produces the very high metallicity systems. Vertical tickmarks show the median metallicity for each sample. {\sc Simba} has a median metallicity of -1.74 that is more consistent with the observed median of -1.67 than the median predicted by the {\sc TD} which is -1.38. This is probably due to the fact that the under-production of high metallicity systems in {\sc Simba} balances the existence of the extremely low-metallicity systems, resulting in a median that is consistent with observations. Given the metallicity evolution, we expect that our simulations will be able to produce more high metallicity systems at lower redshifts when more massive halos are formed. Assuming that the number of low metallicity systems in both simulations is small, {\sc Simba} is in a good agreement with the observed metallicity distribution, whereas {\sc TD} is skewed more towards high metallicity systems. This is largely due to the feedback effects as seen in the last panel in Figure~\ref{fig:gas_dist}. The $\sim$ 3 $\times$ faster wind speed adopted by {\sc TD} pushes metals to larger distances, where most of DLAs form ( n$_{\rm H}$ = 0.01 $-$ 1 cm$^{-3}$) and further contribute to their metal enrichment.

The existence of the very low metallicity systems in {\sc Simba} might be a consequence of the star formation model. To investigate whether the H$_{2}$-regulated star formation model implemented in {\sc Simba} pushes DLAs to have very low metallicity values, we vary the metallicity floor ($Z_{\rm floor}$) which is the initial seed metallicity necessary to switch on star formation in this model. Since the used {\sc Simba} run for these comparisons adopts $\log Z_{\rm floor}=-2$, we run 12.5 h$^{-1}$ Mpc volume of {\sc Simba} with 2$\times$128$^{3}$ dark matter and gas particles each, with three different values of the metallicity floor $\log Z_{\rm floor}$=-1,-2, and -3. We now show the impact of changing $Z_{\rm floor}$ on the DLA metallicites in the right panel of Figure~\ref{fig:MH}. The dark-blue, blue, and steel-blue are the runs with  $\log Z_{\rm floor}$=-1,-2, and -3, respectively, and the vertical tickmarks show the corresponding median metallicity for each run. It is evident that the DLA metallicities overall increase with increasing $Z_{\rm floor}$ as shown by the median vertical tockmarks, each of which increases by approximately 0.5 dex for one order of magnitude increase in $Z_{\rm floor}$. The low metallicity DLAs start to disappear at high values of $Z_{\rm floor}=0.1$, but not completely. This indicates that the feedback effects partially contribute to the low metallicity in DLAs. The increase in $Z_{\rm floor}$ slightly increases the DLA abundance, and we find that a very high value for the $Z_{\rm floor} > 1$ is required to match the observed DLA abundance.  We note that changing the $Z_{\rm floor}$ value has a minimal impact on the stellar mass function. However, a value of $\log Z_{\rm floor}=-2$ is already high since $\log Z_{\rm floor}=-3$ is the commonly used value in these molecular hydrogen SFR models~\citep[e.g.][]{Kuhlen:2013} as motivated by~\citet{Wise:2012} numerical simulations that follow the transition from Population III to Population II star formation.
This comparison indicates that the tail of extremely low-metallicity DLAs in {\sc Simba} is at least partly an artefact of the H$_{2}$-regulated SFR model. Alternatively, if extremely low-$Z$ DLAs exist, then they may be the ancestors of ultra-faint dwarf galaxies~\citep{Cooke:2017}, in which case
{\sc Simba} enables study of their kinematics and nature. We leave these questions for future work.

\section{Conclusions}\label{sec:conc}
We have examined the DLA properties in two state-of-the-art cosmological hydrodynamic simulations: {\sc Simba} and {\sc Technicolor Dawn}. Starting from the same initial conditions, our two simulations were each run down to $z=3$. We have generated mock DLA profiles and their associated metal lines in the redshift range $z=3-5$. The simulations adopt different recipes to form stars, implement galactic feedback, and treat the UVB as summarized in Table~\ref{tab:sims}.

Our two key findings are summarized as follows:
\begin{itemize}
    \item {\sc Simba} under-predicts the observed DLA abundance by a factor of $\sim$ 2, whereas {\sc TD} is more consistent with the measurements (see right panel in Figure~\ref{fig:all}), particularly when post-processing corrections to the UVB amplitude are taken into account (see Figure~\ref{fig:uvb_effs}). This under-production of DLAs is largely due to the {\sc Simba}'s weak feedback effects as compared to {\sc TD} (see Table~\ref{tab:sims}), which in turn boosts the star formation (see left panel in Figure.~\ref{fig:SMF}) and suppresses the DLA incidence rate. Similar trends are seen in the column density distribution function (see left panel in Figure~\ref{fig:all}), except {\sc Simba} resolves much higher column density DLAs than {\sc TD}, as {\sc Simba} continues to suppress feedback in massive galaxies. This results in a good agreement for both simulations with the observed \HI\ density (see middle panel in Figure~\ref{fig:all}).
    
    \item {\sc Simba} is more consistent with the observed DLA metallicity distribution, whereas {\sc TD} is skewed towards high metallicity systems (see left panel in Figure~\ref{fig:MH}). {\sc Simba} further predicts a population of DLAs with metallicities much lower than any observed to date~\citep[e.g.][]{Cooke:2017}. This population is sensitive to the details of the H$_{2}$-regulated SFR model (see right panel Figure~\ref{fig:MH}). Both simulations agree with observed slope of DLA metallicity evolution with redshift (see middle panel in Figure~\ref{fig:MH}).
\end{itemize}

Our comparisons are entirely limited to the simulation resolution and dynamic range. More DLAs are usually found in higher resolution set-up, such as in zoom-in simulations~\citep[see][]{Rhodin:2019}. The unique aspect in this study is the intrinsic difference between these simulations in the star formation models and the inhomogeneous UVB treatment. This work sets the stage for more interesting inquiries on the use of DLAs to constrain galaxy formation models. Future inquiries will include:
\begin{itemize}
    \item Exploring DLA kinematics in relation to the hosing properties between both simulations. 
    \item Studying the nature of metal poor DLAs in connection with ultra-faint dwarf galaxies at high redshift.
    \item  Stuyding the nature of dusty DLAs at high column densities, and the effect of dust bias in DLA selection.
\end{itemize}

Our results have already shown that how DLA observations can play a key role to constraining the star formation recipes and feedback effects in galaxy formation models.

\section*{Acknowledgements}
The authors acknowledge helpful discussions with Joseph Burchett, John Chisholm, David Chih-Yuen Koo and Neal Katz. We particularly thank the referee, Simeon Bird, for his constructive comments which have improved the paper quality significantly. Simulations and analysis were performed at UWC's {\sc Pumbaa}, IDIA/{\sc Ilifu} cloud computing facilities and NMSU's {\sc DISCOVERY} supercomputer. This work also used the Extreme Science and Engineering Discovery Environment (XSEDE), which is supported by National Science Foundation grant number ACI-1548562, and computational resources (Bridges) provided through the allocation AST190003P. 
RD acknowledges support from the Wolfson Research Merit Award program of the U.K. Royal Society.
This work used the DiRAC@Durham facility managed by the Institute for Computational Cosmology on behalf of the STFC DiRAC HPC Facility. The equipment was funded by BEIS capital funding via STFC capital grants ST/P002293/1, ST/R002371/1 and ST/S002502/1, Durham University and STFC operations grant ST/R000832/1. DiRAC is part of the National e-Infrastructure.

\bibliographystyle{mnras}
\bibliography{DLA.bib}
\bsp	% typesetting comment
\label{lastpage}
\end{document}